\documentclass[aps,prb,twocolumn,superscriptaddress,showpacs,amsmath,amssymb]{revtex4}

\usepackage{graphicx}
\usepackage{dcolumn}
\usepackage{booktabs}
\usepackage{bm}
\usepackage{hyperref}
\usepackage{color}
\usepackage{multirow}

\newcommand{\cpT}{\ensuremath{c_{p}/T}}
\newcommand{\cp}{\ensuremath{c_{p}}}
\newcommand{\cel}{\ensuremath{c_{\rm el}}}
\newcommand{\cph}{\ensuremath{c_{\rm ph}}}
\newcommand{\kel}{\ensuremath{\kappa_{\rm el}}}
\newcommand{\kph}{\ensuremath{\kappa_{\rm ph}}}
\newcommand{\ktot}{\ensuremath{\kappa_{\rm tot}}}

\newcommand{\WTST}{W$_{1-x}$Ta$_x$Se$_{1.6}$Te$_{0.4}$}
\newcommand{\WTS}{W$_{1-x}$Ta$_x$Se$_{2}$}

\begin{document}

\title{Modification of electronic structure and thermoelectric properties of hole-doped tungsten dichalcogenides}

\author{M.~Kriener}
\email[]{markus.kriener@riken.jp}
\author{A.~Kikkawa}
\author{T.~Suzuki}
\author{R.~Akashi}
\altaffiliation[present address: ] {Department of Physics, The University of Tokyo, Tokyo 113-0033, Japan}
\author{R.~Arita}
\affiliation{RIKEN Center for Emergent Matter Science (CEMS), Wako, Saitama 351-0198, Japan}
\author{Y.~Tokura}
\affiliation{RIKEN Center for Emergent Matter Science (CEMS), Wako, Saitama 351-0198, Japan}
\affiliation{Department of Applied Physics and Quantum-Phase Electronics Center (QPEC), University of Tokyo, Hongo, Tokyo 113-8656, Japan}
\author{Y.~Taguchi}
\affiliation{RIKEN Center for Emergent Matter Science (CEMS), Wako, Saitama 351-0198, Japan}

\date{\today}

\begin{abstract}
We present a study on the modification of the electronic structure and hole-doping effect for the layered dichalcogenide WSe$_2$ with a multi-valley band structure, where Ta is doped on the W site along with a partial substitution of Te for its lighter counterpart Se. By means of band-structure calculations and specific-heat measurements, the introduction of Te is theoretically and experimentally found to change the electronic states in WSe$_2$. While in WSe$_2$ the valence-band maximum is located at the $\Gamma$ point, the introduction of Te raises the bands at the $K$ point with respect to the $\Gamma$ point. In addition, thermal-transport measurements reveal a smaller thermal conductivity at room temperature of \WTST\  than reported for \WTS. However, when approaching 900~K, the thermal conductivities of both systems converge while the resistivity in \WTST\ is larger than in \WTS, leading to comparable but slightly smaller values of the figure of merit in \WTST.
\end{abstract}


\maketitle

\section{Introduction}
Bulk dichalcogenides $TX_2$ with transition metals $T$ and $X=$~S, Se, or Te belong to the class of layered semimetals / semiconductors\cite{wilson69a} which attracted considerable attention for decades. They exhibit a rich variety of physical properties, such as superconductivity in, e.g., NbSe$_2$ or TaS$_2$,\cite{revolinsky65a,vanmaaren67a} coexisting but competing charge-density wave order,\cite{wilson74a,morris75a,wilson75a,castroneto01a,valla04a} and  thermoelectrical device functions.\cite{brixner62a} In many cases, these materials possess Fermi surfaces consisting of many valleys, which are somehow related to, or responsible for, these phenomena. 

The discovery of graphene has further increased the interest in layered materials with valley degrees of freedom. It was found that the physical properties of bulk layered materials can be significantly altered when thinning them down to atomically flat layers.\cite{novosolev05a,novosolev05b,yzhang05a} Naturally, different layered systems came into the focus of research. Among them are the monolayered dichalcogenides which are often regarded as two-dimensional semiconducting or semimetallic analogs to graphene.\cite{qzhang14a,behnia12a,gwang14a} The electronic band structure plays a crucial role also in the physics of these thin materials with impact on spintronics,\cite{ochoa13a} optoelectronics devices,\cite{ross14a,baugher14a,pospischil14a} or the emerging field of valleytronics, e.g., in MoS$_2$, MoSe$_2$, WS$_2$, and WSe$_2$. \cite{dxiao07a,behnia12a,qhwang12a,dxiao12a,rycerz07a,zzhu12a}
The strong spin-orbit interaction in some transition-metal dichalcogenides and the lack of inversion symmetry in their monolayer variants lead to valley-spin coupling, adding a new feature to their zoo of exotic properties.\cite{dxiao12a,zzhu12a,ochoa13a,hyuan13a,hzlu13a,wyshan13a} Another example of their intriguing nature is the electric-field-induced superconductivity with an optimum $T_{\rm c}$ exceeding 10~K in an electric double-layer transistor structure made from MoS$_2$.\cite{taniguchi12a,jye12a,yzhang12a,radisavijevic13a} The band structures of the aforementioned hexagonal multilayer (bulk) and monolayer WS$_2$, WSe$_2$, and MoS$_2$ differ as well. While the bulk systems exhibit indirect band gaps, the monolayered counterparts have direct band gaps at the inequivalent $K$ and $K^\prime$ points at the corners of the hexagonal Brillouin zone.\cite{kfmak10a,splendiani10a,wzhao13a,ochoa13a,gwang14a} Therefore, to control the band structure is one of the most important issues in the study of those $TX_2$ compounds.

The presence of many valleys in the band structure is in general interesting also for thermoelectrical applications since it is well known to enhance the thermoelectrical performance.\cite{rowe95} Thermoelectrical materials are also in the focus of current research because they offer the possibility to transform thermal waste heat back into usable electrical power.\cite{rowe78a,mahan97a,snyder08a,heremans12a,pei12a} A measure of the thermoelectric efficiency of such materials is the figure of merit (FOM) $ZT = S^2 T/(\rho\kappa)$. It consists of the thermopower or Seebeck coefficient $S$, the longitudinal resistivity $\rho$, and the total thermal conductivity $\kappa= \kel+\kph $, where \kel\ and \kph\ are the contributions of the mobile charge carriers and the lattice, respectively. The term $S^2/\rho$ is often referred to as the power factor.  To maximize $ZT$, a large thermopower and a small resistivity along with a small thermal conductivity are required. Conventional metals are usually not good candidates for high-efficiency thermoelectrics since the Seebeck coefficient is generally small due to the Fermi degeneracy. Instead, doped semiconductors are promising materials where it is possible to control the charge-carrier concentration and hence the electrical conductivity and the Seebeck coefficient by doping. However, most of such systems suffer from a good lattice heat conductivity. The guiding principle can be abbreviated as ``phonon glass + electron crystal'',\cite{snyder08a,rowe95} i.e., a system which ideally consists of independent charge- and heat-transport channels to obtain low $\rho$ and $\kappa$ values at the same time. In this context, bulk dichalcogenides attracted much interest when it was found that several of them exhibit at room temperature a small heat conductivity of only $\sim 2$~W/K~m, among them WSe$_2$.\cite{brixner62a} Back in the 1960s and 1970s, this material was intensively studied by doping into both the W (cation) and the Se (anion) lattice sites\cite{brixner63a,brixner63b,hicks64a,revolinsky64a,champion65a,mentzen76a} as well as by intercalation of metal elements to bridge the chalcogen layers.\cite{whittingham75a}

WSe$_2$ is a $p$-type semiconductor and crystallizes in the hexagonal $P6_3/mmc$ structure (space group 194), usually abbreviated as 2H-WSe$_2$.\cite{glemser48a,brixner62a} It consists of trilayer building blocks Se\,--\,W\,--\,Se with strong covalent bonds. These blocks are separated by only weakly bonded van der Waals gaps. Each W ion is coordinated by six Se ions with a trigonal prism configuration. The unit cell consists of two formula units along the $c$ axis. Several band-structure calculations can be found in literature, the most recent one in Ref.~\onlinecite{hyuan13a}. The valence-band maximum lies at the $\Gamma$ point (zone center) almost degenerate with the only slightly lower-lying band at the $K$ (and $K^\prime$) point. While the band dispersion at the $\Gamma$ point has an almost isotropic nature, at the $K$ point it is anisotropic. By doping pentavalent Ta$^{5+}$ on the W$^{4+}$ site, holes are introduced into the valence band, and the large electrical resistivity of pure WSe$_2$ is successfully suppressed: \WTS\ becomes metallic at small Ta concentrations $x\approx 0.03$ with room-temperature resistivities of the order $\sim$ m$\Omega$cm. Up to $x\approx 0.35$, the structure remains hexagonal $P6_3/mmc$ with $p$-type conduction.\cite{brixner62a, brixner63a,hicks64a} On the other hand, there are fewer works published on WSe$_{2-y}$Te$_y$, i.e., the substitution of isovalent Te$^{2-}$ for Se$^{2-}$. The end member WTe$_2$, crystallizing in an orthorhombic structure ($Pmmm$), is a semimetal with a negative Hall coefficient. It seems that at least up to $y=0.5$ the hexagonal WSe$_2$ structure is retained.\cite{champion65a,wilson69a} As for the preparation of these dichalcogenides, it should be noted that they do not melt congruently and dissociate at elevated temperatures. This was discussed earlier as a problem in achieving samples with high packing densities.\cite{brixner63a} Also, the physical properties were not very reproducible.\cite{brixner63a}

Here we present a comprehensive study on \WTST\ with $0\leq x \leq 0.06$ by means of band-structure calculations, transport, and thermodynamical measurements to elucidate the effect of Te doping on the electronic structure. Interestingly, we found evidence that the valence-band maximum shifts from the $\Gamma$ point in WSe$_2$ toward the $K$ point when substituting Se by Te while keeping the crystal structure, which is reminiscent of the aforementioned situation in the monolayer dichalcogenides. The introduction of Te was also found to further suppress the thermal conductivity around room temperature. 

The paper is organized as follows: In the next section (Sec.~\ref{prepmeth}), the sample preparation and experimental and theoretical calculation methods are described. After overviewing the electronic band structure (Sec.~\ref{bandstr}), the basic transport properties and structural data below room temperature for \WTST\ are introduced, together with those for \WTS\ for comparison (Sec.~\ref{sampchar}). Next (Sec.~\ref{specheat}) we focus on the change in the electronic states of WSe$_2$ due to the Se replacement with Te by discussing the results of specific-heat data in the light of the band-structure calculations. In the latter half of this paper (Sec.~\ref{thermprop}), we present the thermoelectric properties of \WTST, i.e., thermal-transport data up to $\sim 850$~K from which the FOM is calculated. Section \ref{summ} is devoted to the summary of the present work.

\section{Sample Preparation and Methods}
\label{prepmeth}
Polycrystals\cite{SingleCrystcomment} of \WTST\ for $x=0$, 0.02, 0.025, 0.03, 0.035, 0.0375, 0.04, and 0.06 were prepared in three steps. First, stoichiometric amounts of the elements W (purity 99.9\%), Ta (99.9\%), Se (99.999\%), and Te (99.999\%) were thoroughly ground, mixed, sealed into evacuated quartz tubes, and kept for 48 h at $700^{\circ}$C. The resulting reaction product was free-flowing blackish powder. Second, the powder was reground, pelletized, again sealed into evacuated quartz tubes, and kept for 72~h to 96~h at $1000^{\circ}$C. Third, the reaction product was ground again. Approximately 600~mg of the powder was used to synthesize a final batch by employing a high-pressure ($p$) / high-temperature ($T$) technique using a cubic anvil cell. The latter approach was chosen to overcome the aforementioned problem to achieve high-packing densities.\cite{brixner63a} Practically, the powder was first pressed at 2 GPa at room temperature. Next, the temperature was increased to about $1100^{\circ}$C. This temperature was kept for about 10 minutes, then the material was thermally quenched. The pressure was released after the temperature had dropped back to room temperature. For each of the three steps, the reaction product was checked by x-ray diffraction (XRD). The targeted compounds had already formed after the first step although the respective XRD peaks were very broad. This probably indicates a disordered stacking of the characteristic (Se,Te)\,--\,W\,--\,(Se,Te) trilayers. The second and third reaction steps respectively lead to a sharpening of the XRD patterns. For the various measurements, rectangularly- and cylindrically-shaped samples were cut from these batches.\cite{samplecomment} We note that we did not observe any cleavage-like behavior when cutting the samples. This is probably a result of the application of hydrostatic pressure in our cubic anvil press, leading to more isotropic samples without preferred orientation in spite of the layered structure of WSe$_2$. As a control experiment, we also prepared pristine WSe$_2$ and  \WTS\ ($x = 0.02$, 0.03, 0.04, 0.05, 0.06)  by the same high-$p$ / high-$T$ synthesis but without any prereaction, yielding sharp XRD line patterns. Comparably sharp line patterns for \WTST\ were only achieved after the three-step growth process. We note that some batches contained small amounts of unreacted Se / Te after the high-$p$ / high-$T$ synthesis step. From TG-DTA (Rigaku ThermoPlus Evo TG 8120) analyses, we found that the samples lost some weight when heating them up. At higher temperatures the system starts to dissociate in agreement with an old report.\cite{brixner63a} Therefore, we restricted our high-temperature experiments to below approximately 850~K. Moreover, most of the examined specimens were annealed 24~h at $150^{\circ}$C before the high-$T$ measurements. 

Below 300~K, the longitudinal resistivity $\rho_{xx}$ and the Hall resistivity $\rho_{yx}$ were measured by a standard five-probe technique using a commercial system (Quantum Design, PPMS). Low-$T$ longitudinal Seebeck $S_{xx}$ and thermal conductivity $\kappa_{xx}$ data\cite{Skappacomment} were taken with two separate home-built setups, each mounted on a PPMS cryostat. Above room temperature up to approximately 850~K, $\rho_{xx}$ and $S$ were measured simultaneously in a ZEM-3 system (ULVAC Technologies), where the sample is held in He atmosphere by two Pt or Ni electrode stamps acting as current leads. Two thermocouples were pressed to one sample surface acting as voltage pads and were used for the Seebeck measurement. High-$T$ resistivity and thermopower data were taken upon cooling. The respective low-$T$ measurements were carried out using the same samples after the high-$T$ experiment. The thermal conductivity above room temperature was estimated using the formula $\kappa = D \,\cp\,d_{_{\rm 300K}}$ with the thermal diffusivity $D$, the room-temperature density of the respective samples $d_{_{\rm 300K}}$, and the specific heat \cp. The thermal diffusivity was measured by employing the laser-flash method in a commercial Netzsch LFA-457 apparatus. The density of the samples was estimated from their mass and dimensions. Specific-heat data below 300~K were measured by a relaxation-time method using the PPMS. The addenda heat capacity was measured before mounting the sample and eventually subtracted from the total signal. Since the specific heat of \WTST\ at room temperature already exceeds 95\% of the classical Dulong-Petit limit $c_{_{\rm DP}}$, \cp\ at higher temperatures was calculated as $\cp = \gamma T + c_{_{\rm DP}}$ with the Sommerfeld parameter $\gamma$, which was determined from \cp\ vs $T$ plots at low temperatures.\cite{cpcvcomment}

The first-principles band-structure calculations were performed with the WIEN2k code employing the full-potential linearized augmented plane-wave method.\cite{blaha01} We used the Perdew-Burke-Ernzerhof exchange-correlation functional.\cite{perdew96a} In the calculations for WSe$_{2}$, the measured lattice parameters ($a=3.289$~\AA, $c=12.988$~\AA) and the positional parameter  $z=0.129$ (W\,--\,Se layer distance) reported in Ref.~\onlinecite{coehoorn87a} were employed. For hypothetical 2H-WTe$_{2}$, we optimized the parameters by using the scalar-relativistic approximation\cite{koelling77a} ($a=3.555$~\AA, $c=14.447$~\AA, and $z=0.126$). Spin-orbit coupling was also taken into account as a relativistic effect in the band-structure calculations.\cite{kunes01a} The structure optimization (band-structure calculation) was carried out with the cutoff $RK_{\rm max}=8.5$ ($10.0$) and $6 \times 6 \times 4$ ($12 \times 12 \times 8$)~$k$ points. The density of states was calculated with $36 \times 36 \times 8$~$k$ points using the tetrahedron method.\cite{bloechl94a}
 
 \section{Results and Discussion}
 \subsection{Band Structure}
 \label{bandstr}
\begin{figure}[t]
\centering
\includegraphics[width=8.5cm,clip]{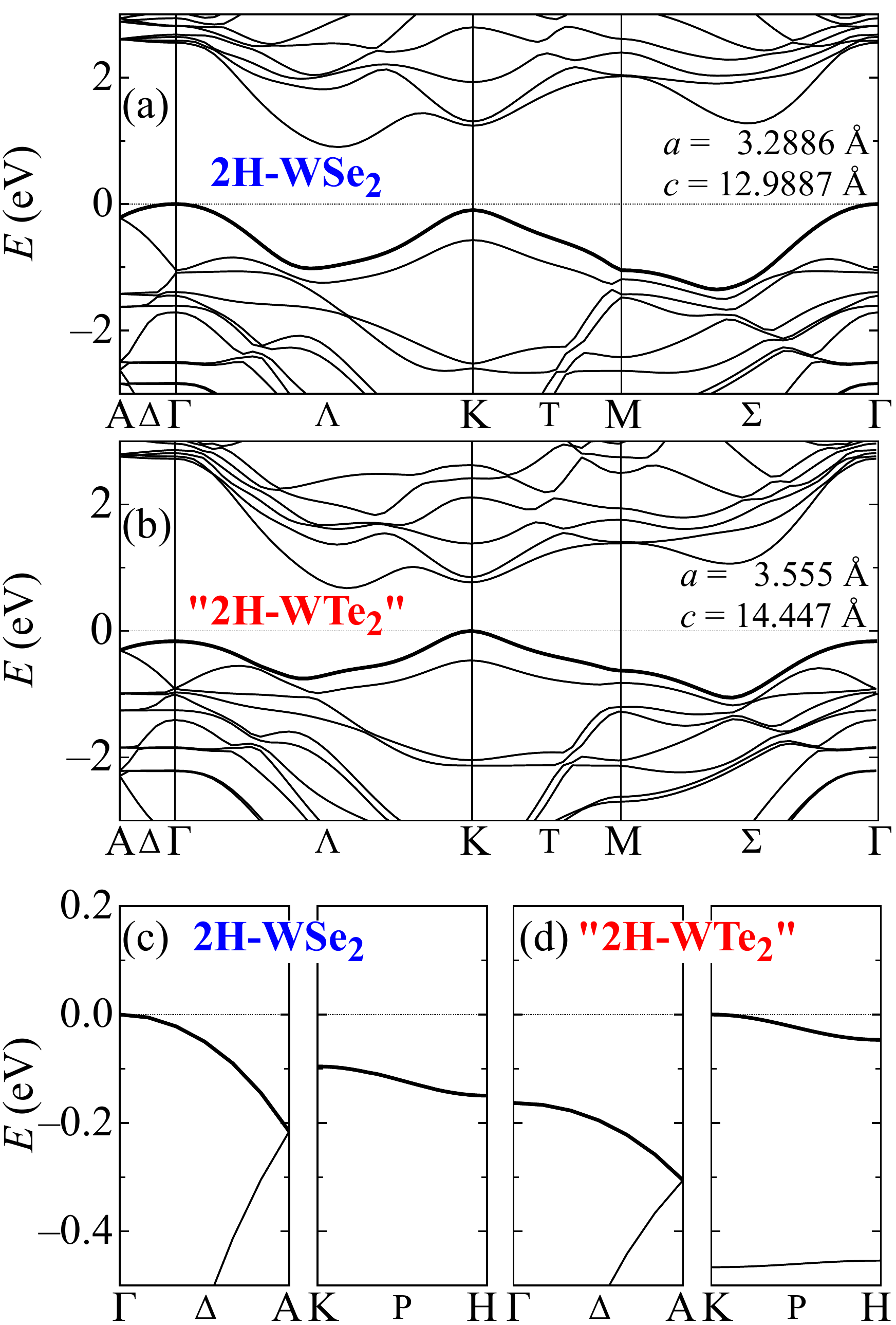}
\caption{(Color online) Band structure calculated for (a), (c) 2H-WSe$_2$ and (b), (d) hypothetical 2H-WTe$_2$. The values of the lattice constants used for the calculations are indicated in panels (a) and (b). For the details, see text. Panels (c) and (d) provide expanded views of the band dispersions along the $k_z$ directions from the $\Gamma$ and $K$ points of the hexagonal Brillouin zone. In each panel, the Fermi level is indicated by a horizontal line ($E=0$~eV).}
\label{fig1}
\end{figure}
Figure~\ref{fig1} summarizes the results of band-structure calculations. In Fig.~\ref{fig1} (a), the band structure based on experimentally determined lattice constants for 2H-WSe$_2$ is plotted. The valence-band maximum is located at the $\Gamma$ point which has an almost isotropic dispersion.\cite{BScomment} The second, almost degenerate valence-band maximum is found at the $K$ and $K^\prime$ points and lies approximately 100~meV lower in energy, yielding a unique band structure as discussed in Ref.~\onlinecite{hyuan13a}. Panel (c) gives an expanded view of the band dispersion along the $k_z$ direction at the $\Gamma$ and $K$ points, i.e., $\Gamma$-$A$ and $K$-$H$, respectively. The in-plane and out-of-plane band widths around the $K$ and $K^\prime$ points are much more different than around the $\Gamma$ point. Hence, the band dispersion around $K$ or  $K^\prime$ has an anisotropic, two-dimensional nature. 

To investigate how the WSe$_2$ band structure changes upon Te doping, we calculated the energy dispersion for hypothetical 2H-WTe$_2$ based on optimized lattice and positional parameters. It is important to mention that hexagonal 2H-WTe$_2$ \textit{does not} exist in nature. In reality, WTe$_2$ crystallizes in the orthorhombic $Pmmm$ structure.\cite{brown66b,ali14b} The result is shown in Fig.~\ref{fig1} (b). Expanded views along the $k_z$ directions from the $\Gamma$ and $K$ points are plotted in panel (d). Interestingly, the substitution of the heavier Te for Se shifts the valence-band maximum from $\Gamma$ to $K$ (and $K^\prime$), but the energy difference between them remains small. Most importantly, this shift has implications on the density of states (DOS) at the onset of the hole band. While the band-edge DOS in 2H-WSe$_2$ is dominated by the almost isotropic band at the $\Gamma$ point, it originates from the more anisotropic bands at the $K$ and $K^\prime$ points in hypothetical 2H-WTe$_2$. As long as the 2H-type hexagonal structure is retained, it is reasonably expected that the relative position of the band maximum gradually and continuously changes upon increasing the Te concentration in WSe$_{2-y}$Te$_y$. The region dominating the band-edge DOS changes from around the $\Gamma$ point to around the $K$ and $K^\prime$ points, when the energy level at the local maxima for the $\Gamma$ and $K$ points become the same. Following our calculations this happens at $y \approx 0.74$.

\subsection{Sample Characterization}
\label{sampchar}
\begin{figure}[t]
\centering
\includegraphics[width=8cm,clip]{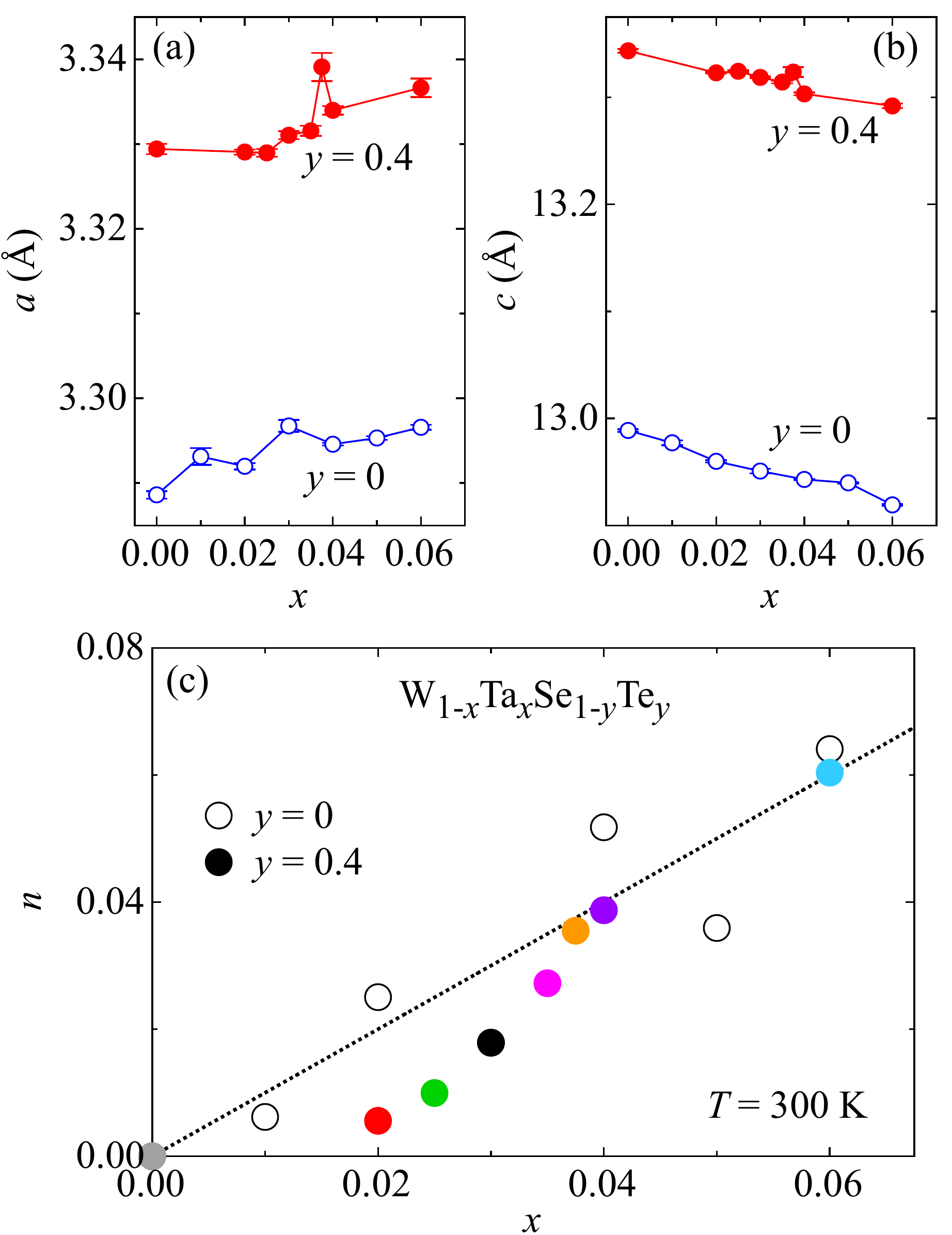}
\caption{(Color online) (a), (b) Evolution of the lattice constants $a$ and $c$ with $x$ for \WTS\  (open blue symbols; $y=0$) and \WTST\  (closed red symbols; $y=0.4$). (c) Actual charge carrier concentration per W site $n$ as estimated from Hall resistivity $\rho_{yx}$ at 300~K is plotted against the nominal Ta-doping concentration $x$ of \WTST\ (filled symbols). Data for \WTS\ (open symbols) are included for comparison; see text. The dotted line indicates $n = x$. The color code of the different data points is the same as in Fig.~\ref{fig4}.} 
\label{fig2}
\end{figure}

Figures~\ref{fig2} (a) and (b) show the Ta-doping dependence of the lattice constants $a$ and $c$, respectively, for \WTST\ ($y=0.4$; filled red symbols) and \WTS\ ($y=0$; open blue symbols). According to the introduction of Ta and Te in the two different lattice sites, the lattice constants change systematically with $x$ and $y$, indicating the successful formation of solid solutions. Ta doping increases the $a$-axis length whereas the $c$ axis shrinks. The increase in the absolute values between $y = 0$ and $y = 0.4$ reflects the replacement of 20\% of the Se$^{2-}$ ions with larger Te$^{2-}$ ions. Both of the Te and Ta ions readily substitute their counterparts. 

The charge-carrier concentration as estimated from Hall resistivity $\rho_{yx}$ measurements at 300~K increases from $\sim 8.7 \times 10^{19}$~cm$^{-3}$ for $x=0.02$ to $\sim 9.4 \times 10^{20}$~cm$^{-3}$ for $x = 0.06$. These data are plotted with filled symbols as charge-carrier concentration per W site $n$ in Fig.~\ref{fig2} (c); $n=0.01$ corresponds to a charge-carrier concentration of $\sim 1.65\times10^{20}$~cm$^{-3}$. We also estimated $n$ at 5~K (not shown), finding only small changes from the values at 300~K. At both temperatures, the Hall resistivity $\rho_{yx}$ is linear in the magnetic field, indicating that there are only hole-type charge carriers. We are not able to measure the Hall resistivity at elevated temperatures, but for low-doped \WTS, a temperature-independent charge-carrier concentration up to $T\sim 900$~K was reported earlier.\cite{hicks64a} For comparison, respective data for \WTS\ are also shown in open symbols. While the actual charge-carrier concentrations $n$ are close to the nominal doping levels $x$ for \WTS, this is not the case for $x < 0.04$ in the \WTST\ system. We find a systematic suppression of $n$ compared to the nominal carrier concentration corresponding to the Ta-doping level $x$, i.e., one hole per Ta$^{5+}$. Possibly, some of the introduced charge carriers are annihilated in the low-doped samples due to compensation effects arising from deficiency in the Se\,--\,Te stoichiometry. Such an effect is absent in the ``pure'' diselenide system. This annihilation does not seem to be active when a certain amount of Ta is doped and the system becomes metallic. In Ref.~\onlinecite{hicks64a}, the author reports a smaller than nominal charge-carrier concentration also for \WTS\ and speculates about a compensation effect due to impurities at low doping concentrations. 

\begin{figure}[t]
\centering
\includegraphics[width=7.5cm,clip]{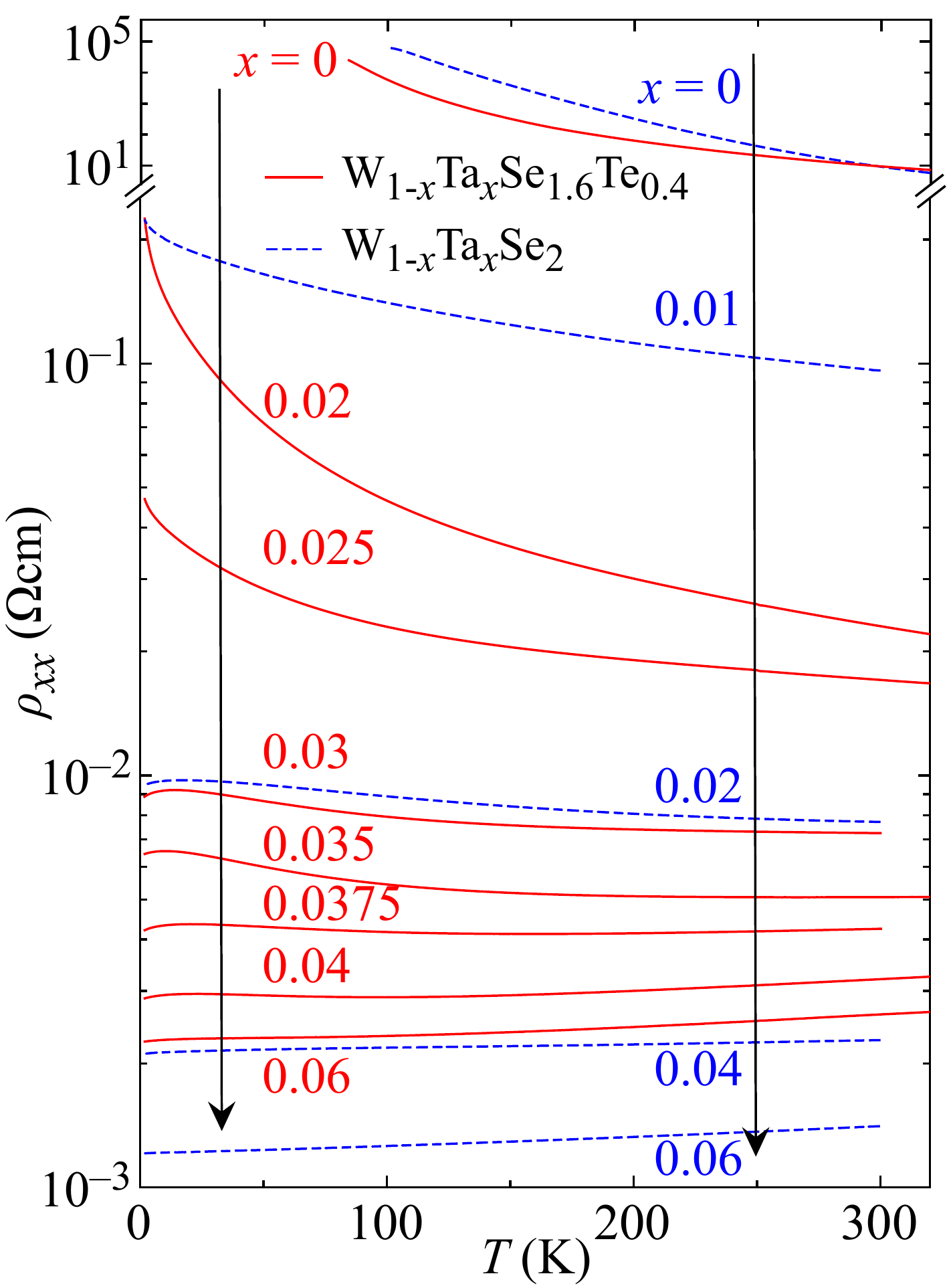}
\caption{(Color online) Temperature dependence of the resistivity $\rho_{xx}$ below room temperature of \WTS\ (dashed blue lines; $y=0$) and \WTST\ (solid red lines; $y = 0.4$) for $0\leq x \leq 0.06$. Note the change in the scale of the vertical axis.}
\label{fig3}
\end{figure}
Figure~\ref{fig3} summarizes the resistivity $\rho_{xx}$ data below 300~K for both the Se-\textit{unsubstituted} \WTS\ (dashed blue lines) and the Se-\textit{substituted} \WTST\ (solid red lines). In both systems, the partial replacement of W$^{4+}$ by Ta$^{5+}$ introduces holes and systematically lowers the resistivity with $x$. Already a small Ta concentration of $x \leq 0.02$ drastically suppresses $\rho_{xx}$ by orders of magnitude compared to $x=0$.\cite{brixner63a,hicks64a} WSe$_2$ and WSe$_{1.6}$Te$_{0.4}$ have room-temperature resistivities of the order of  $\sim 10$~$\Omega$cm whereas $\rho$ at 300~K of W$_{0.99}$Ta$_{0.01}$Se$_2$ has dropped to $\sim 0.1$~$\Omega$cm, and W$_{0.98}$Ta$_{0.02}$Se$_2$ exhibits already less than $10^{-2}$~$\Omega$cm.  For \WTST, the resistivity drops for $x=0.03$ below $10^{-2}$~$\Omega$cm. Metallic behavior is observed in both series above $x \gtrsim 0.04$, although the absolute values of $\rho_{xx}$ remain in the m$\Omega$cm range. The resistivity of the doping series \WTS\ (see also Refs.~\onlinecite{brixner63a} and \onlinecite{hicks64a}) drops faster than in \WTST. This is probably due to additional disorder caused by the partial substitution of Te for Se. We note that in Ref.~\onlinecite{brixner63a} especially for $x=0.01$, smaller values are reported for $\rho_{xx}$, whereas those of the metallic compositions are comparable.

\subsection{Specific Heat}
\label{specheat}
\begin{figure}[t]
\centering
\includegraphics[width=8cm,clip]{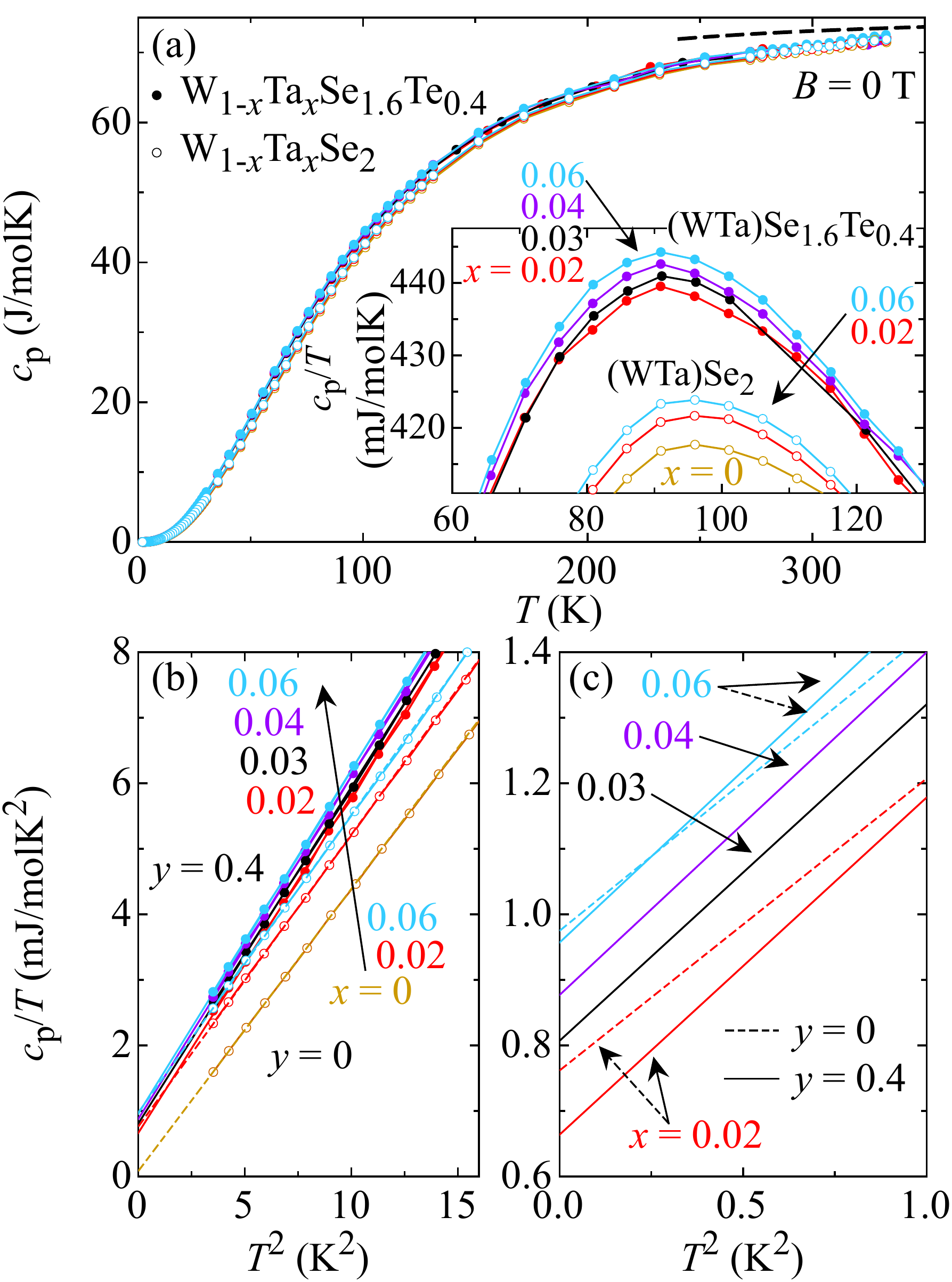}
\caption{(Color online) Specific-heat \cp\ data of \WTST\ (filled symbols) and \WTS\ (open symbols): (a) \cp\ vs $T$ as measured. The dashed line depicts the specific heat calculated in the Debye model plus electronic contribution for $x=0.06$; see text. The inset shows an expanded view around 100\,K displayed as \cpT\ vs $T$. (b) The same data plotted as \cpT\ vs $T^2$ at low $T$. The lines therein are linear fits to the data. An expanded view of the extrapolation to 0~K is shown in (c). Solid lines refer to \WTST\ ($y = 0.4$), dashed lines to \WTS\ ($y = 0$). The intercepts at $T = 0$~K give the electronic specific-heat coefficients $\gamma$; see text.} 
\label{fig4}
\end{figure}
Next, we discuss the Te-doping effect in terms of specific-heat data below $T\sim 330$~K down to approximately 1.9~K. There is no specific-heat study on the doped systems reported in the literature. For the mother compound WSe$_2$, we found only one publication reporting \cp\ data above 60~K.\cite{bolgar90a} Here we show that the analysis of the \cp\ data provides important information about the change in the electronic states when going from \WTS\ to \WTST. These data are summarized in Fig.~\ref{fig4} for \WTST\ ($x=0.02$, 0.03, 0.04, and 0.06; filled symbols) along with data for \WTS\ ($x = 0$, 0.02, and 0.06; open symbols). In Fig.~\ref{fig4} (a), the data are displayed as \cp\ vs.\ $T$. The specific heat of all samples is very similar on this scale. Around room temperature, each sample has released more than $\sim95$\% of the entropy expected in the classical Dulong-Petit limit $c_{_{\rm DP}}$ \textit{plus} its respective electronic contribution $\gamma T$. This is exemplarily indicated as a dashed line in Fig.~\ref{fig4} (a) for $x = 0.06$. 

However, the expanded plot shown as $c_{\rm p}/T$ vs $T$ in the inset of Fig.~\ref{fig4} (a) reveals that there is a clear impact on \cp\ when changing $x$ and $y$. The difference between \cp\ of \WTS\ ($y=0$) and \WTST\ ($y=0.4$) is due to the change in the phonons (lattice) caused by the replacement of 20\,\% Se with Te. Moreover, we could also successfully resolve very small but systematic changes in both systems due to the increase of the Ta concentration $x$, and hence a change of the electronic specific heat. This can be seen in Fig.~\ref{fig4} (b), which shows the specific heat at low temperatures displayed as $\cp/T$ vs $T^2$, where the phononic contribution is small and the electronic sector can be studied. Figure~\ref{fig4} (c) provides an expanded view of the extrapolation of the low-$T$ data to 0~K (linear fits to the respective data). Solid lines refer to \WTST, dashed lines to \WTS. 

Conventional Debye fits to the data \cp\ vs $T$ for $T\leq 5$~K using 
\begin{equation}
\cp = \cel + \cph = \gamma T + A_3 T^3
\end{equation}
yield good descriptions with the electronic specific-heat coefficient $\gamma$ and the coefficient of the phononic part $A_3$. From the latter, the Debye temperatures $\Theta_{\rm D}$ of each sample were calculated via $A_3 = (12/5)\,\pi^4 N N_{\rm A} k_{\rm B}/\Theta_{\rm D}^3$ with the number of atoms per formula unit $N=3$, the Avogadro number $N_{\rm A}$, and the Boltzmann constant $k_{\rm B}$. For all measured samples of the \WTST\ system, we found $\Theta_{\rm D}\approx 224$\,K and for the \WTS\ system, $\Theta_{\rm D}\approx 235$~K for $x=0$, 0.02, 0.06, and 211~K ($x=0.01$), and 228~K ($x=0.04$). The difference in the released entropy when going from the pristine diselenide system ($y=0$) to the Se-substituted system ($y=0.4$) up to approximately 200~K where the different \cp\ data start to converge amounts to approximately 3~J/(mol~K). 
\begin{figure}[t]
\centering
\includegraphics[width=7.5cm,clip]{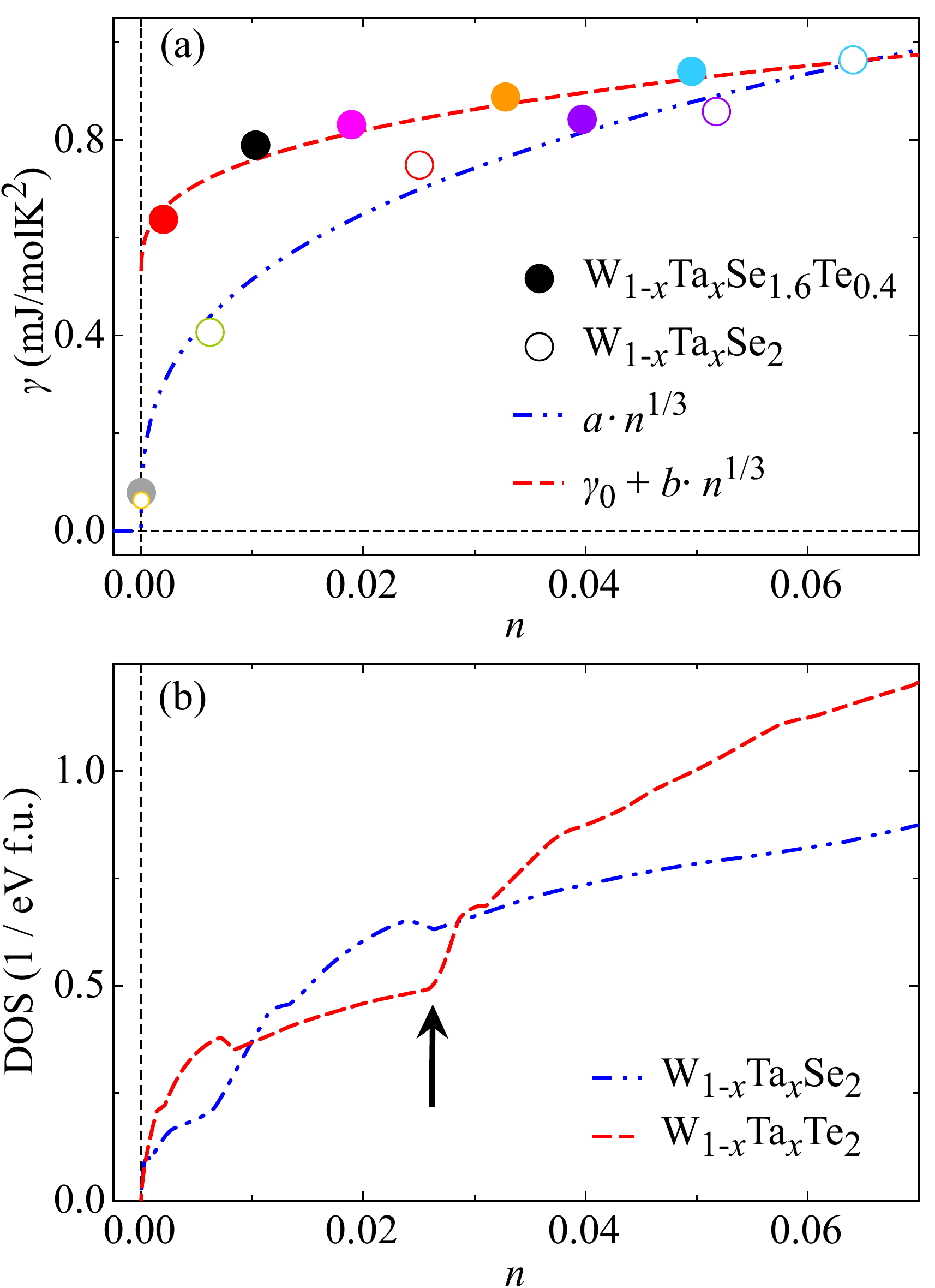}
\caption{(Color online) (a) Charge-carrier concentration dependence of $\gamma$ of \WTST\ (filled symbols) and \WTS\ (open symbols). The dashed (red) and dashed-dotted (blue) lines are fits to the data assuming differences between the two systems in the underlying band structure; see text. The color code of the different data points is the same as in Fig.~\ref{fig4}. Since the data points for both mother compounds without Ta are overlapping on this scale, the one referring to WSe$_2$ is plotted with a smaller symbol. (b) Calculated density of states of \WTS\ and hypothetical 2H-W$_{1-x}$Ta$_x$Te$_2$ as a function of charge-carrier concentration per W site $n$. The arrow marks the onset of filling of the isotropic band at the $\Gamma$ point for hypothetical 2H-W$_{1-x}$Ta$_x$Te$_2$; see text.}
\label{fig5}
\end{figure}

The undoped compounds without Ta exhibit an one-order-of-magnitude smaller electronic contribution to the specific heat as already suggested by the insulating nature of these specimens with a band gap. For both systems we observe a systematic change of $\gamma=(\pi^2 k_{\rm B}^2/3){\rm DOS}$ with the actual charge-carrier concentration $n$ as plotted in Fig.~\ref{fig5}~(a). This reflects the change in the DOS upon hole doping, clearly showing that the filling dependence of $\gamma$ appears to be qualitatively different between the two systems. In view of our band-structure calculations shown in Fig.~\ref{fig1}, the substitution of Te for Se leads to a rise of the bands at the $K$ point with respect to the $\Gamma$ point. Since the band dispersion at the $K$ point is highly anisotropic whereas it is almost isotropic at the $\Gamma$ point, charge carriers in \WTS\ and \WTST\ upon Ta doping may be accommodated into bands of different degree of anisotropy. Motivated by this, we tried to fit the $\gamma(n)$ data points to different formulas: (i) for \WTS\ to 
 \begin{equation}
 \gamma(n)= a\,n^{1/3}
 \end{equation}
 as expected for isotropic bands, and (ii) for \WTST\ to 
 \begin{equation}
 \gamma(n)= \gamma_0 + b\,n^{1/3}.
 \end{equation}
Here, $\gamma_{0}$ is an offset which accounts for the constant DOS of an ideally two-dimensional band structure. These approaches yield the dashed-dotted (\WTS) and dashed (\WTST) lines in Fig.~\ref{fig5}~(a),  well accounting for a qualitative difference between both systems. For \WTST, we find $\gamma_0 = 0.52$~mJ/mol\,K$^2$. We note that for \WTS\ the $\gamma$ value of the undoped compound WSe$_2$ was included to the fit without problems whereas for \WTST, data for $x=0$ had to be excluded to obtain good fitting results, which is another indication for the difference in the DOS of both systems. Since the bands at the $\Gamma$ and $K$ points change their relative position when going from WSe$_2$ to WTe$_2$, our observation is reasonable. Upon increasing the Te concentration, the bands at both points are getting closer in energy: In \WTST, charge carriers also populate the bands of highly anisotropic character at the $K$ point. Hence the constant offset $\gamma_{0}$ is needed to describe $\gamma(n)$ properly. 

To further strengthen this argument, the calculated DOS as a function of charge-carrier concentration per W site $n$ is shown in Fig.~\ref{fig5}~(b). Again the dashed-dotted line refers to \WTS\ and the dashed line to hypothetical 2H-W$_{1-x}$Ta$_x$Te$_2$. The ditelluride system exhibits an initial steplike increase of the DOS as expected for the filling of charge carriers into a highly anisotropic band at the $K$ point. The arrow in Fig.~\ref{fig5}~(b) marks the onset of filling into the isotropic bands at the $\Gamma$ point. By contrast, the diselenide system exhibits a smoother initial filling, indicating the isotropic character of the band structure at the $\Gamma$ point. A direct comparison of Figs.~\ref{fig5}~(a) and (b) is difficult, since \WTST\ contains Se and Te and hence is located in between the two extreme cases shown in panel (b). However, the filling dependence of $\gamma$ for \WTST\ can be well accounted for if the onset of the filling into the isotropic band at the $\Gamma$ point (as indicated by the arrow) nearly coincides with $n=0$.

\subsection{Thermoelectric Properties}
\label{thermprop}
\begin{figure}[t]
\centering
\includegraphics[width=7.5cm,clip]{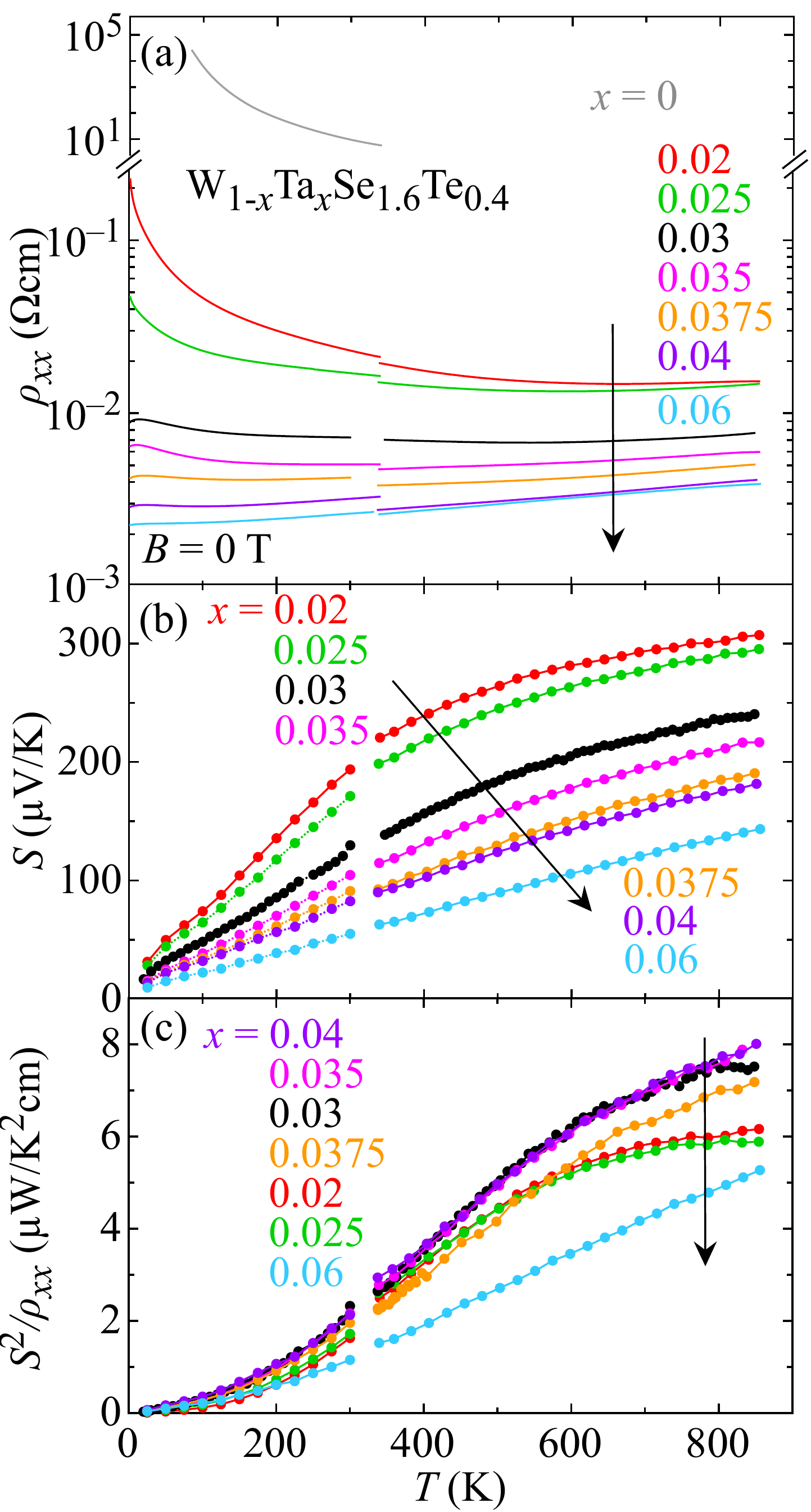}
\caption{(Color online) Transport data of \WTST\ for $0.02 \leq x \leq 0.06$.  (a) Resistivity $\rho_{xx}$, (b) thermopower $S$, and (c) power factor $S^2/\rho_{xx}$ are plotted against temperature up to 850~K. In (a), data for WSe$_{1.6}$Te$_{0.4}$ is also included for comparison.}
\label{fig6}
\end{figure}
\begin{figure}[t]
\centering
\includegraphics[width=7.5cm,clip]{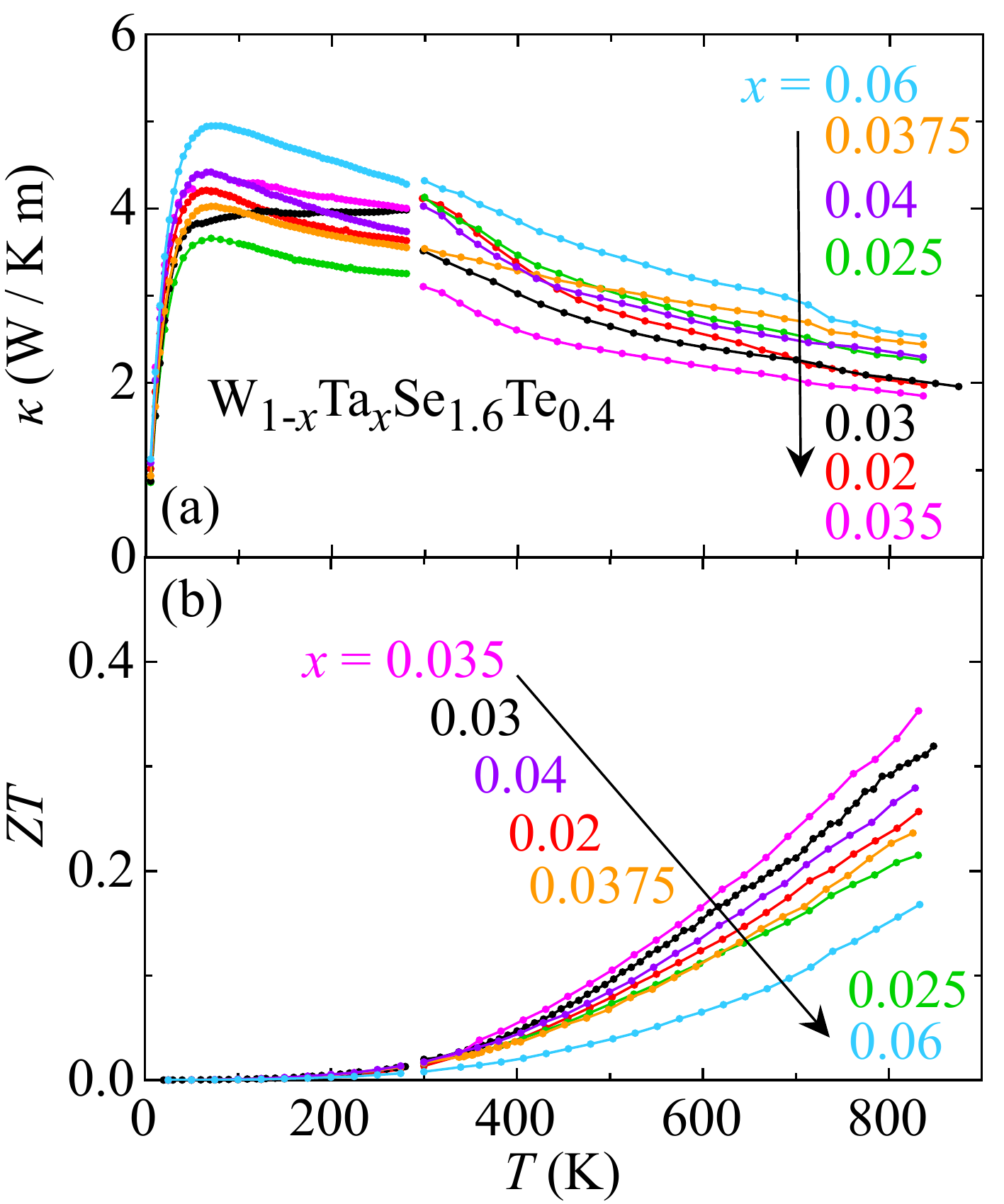}
\caption{(Color online) Temperature dependence of (a) thermal conductivity $\kappa$ and (b) dimensionless thermoelectric figure of merit $ZT$ of \WTST\ below 850~K. }
\label{fig7}
\end{figure}
Having confirmed that the band structure is successfully modified by the partial replacement of Se, next we studied the thermoelectric properties of the \WTST\ system. In Fig.~\ref{fig6}, the temperature dependence of transport data up to $\sim 850$~K for \WTST\ is summarized: (a) resistivity $\rho_{xx}$, (b) thermopower $S$, and (c) the corresponding power factor $S^2/\rho_{xx}$. Although different experimental setups were used for the low-$T$ and high-$T$ measurements, respective resistivity data sets agree with each other around room temperature within the experimental error bars, except for $x = 0.04$. For $x < 0.04$, each sample exhibits a minimum in $\rho_{xx}$ at elevated temperatures. Above about 650~K, all of them feature a metal-like positive slope of $\rho_{xx}$ against $T$. The low-$T$ and high-$T$ Seebeck coefficients shown in panel (b) also agree well around 300~K. The Ta doping leads to a very systematic change of $S$, too. For all $x$, the thermopower increases with $T$ up to the highest measurement temperature 850~K, although the slope flattens above room temperature for most of them. For degenerated semiconductors, the thermopower is expected to be linear in temperature, which we indeed observe below the flattening. The most metallic sample $x = 0.06$ retains the linearity up to the highest measurement temperature. At around 850~K, the largest thermopower $S \approx 300$~$\mu$V/K is found for the low-doped samples. From the data shown in Fig.~\ref{fig6} (a) and (b), the power factors $S^2/\rho_{xx}$ were calculated for each sample and plotted in (c). We observe a maximum in the power factor of $S^2/\rho_{xx} \approx 8$~$\mu$W/K$^2$cm at around 850~K for samples in the intermediate doping range $x=0.03 - 0.04$. This is due to the benefit of an increased conductivity while the thermopower is still fairly large even for the more metallic specimen with $x = 0.04$. 

The thermal conductivity and respective FOM $ZT = S^2 T/(\rho\kappa)$ for these samples are plotted in Figs.~\ref{fig7}~(a) and (b), respectively. For the thermal-conductivity measurements, not only the experimental setups differ between high-$T$ and low-$T$ measurements, but also different samples were used.\cite{samplecomment} The agreement between low-$T$ and high-$T$ data is less satisfactory than in the resistivity and thermopower measurements, but still within the acceptable range. Except for $x=0.03$, all data exhibit a maximum in $\kappa(T)$ below 100~K. Toward higher temperatures, the thermal conductivity for all $x$ continuously decreases due to the three-phonon-scattering process and does not show any signature of saturation even at 850~K. The corresponding FOM plotted in Fig.~\ref{fig7}~(b) increases with increasing temperature for all $x$. Above 800~K a maximum $ZT\approx 0.35$ is found for $x = 0.035$. 

\begin{figure}[t]
\centering
\includegraphics[width=8.5cm,clip]{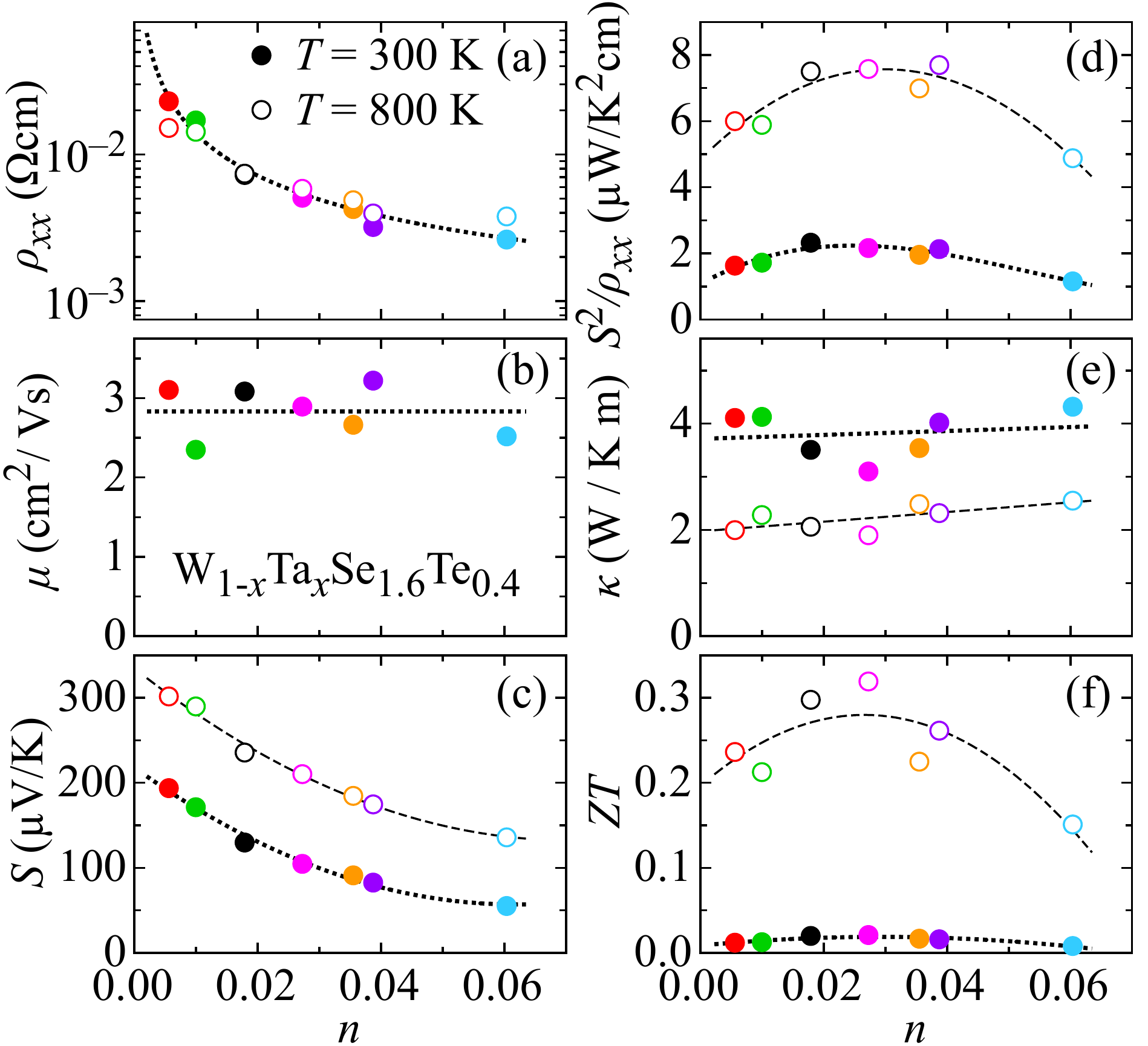}
\caption{(Color online) Summary of transport data for the \WTST\ system at $T=300$~K (filled symbols) and 800~K (open symbols). (a) Resistivity, (b) Hall mobility, (c) thermopower, (d) power factor, (e) thermal conductivity, and (f) FOM are plotted as a function of charge-carrier concentration per W site $n$. The lines in the panels are guides to the eyes. The color code of the different data points is the same as in Fig.~\ref{fig4}.}
\label{fig8}
\end{figure}

For a better comparison, we replotted various properties of  \WTST\ against the actual charge-carrier concentration $n$ at 300~K (filled symbols) and at 800~K (open symbols) in Fig.~\ref{fig8}: (a) $\rho_{xx}$, (b) Hall mobility $\mu = (n e \rho_{xx})^{-1}$ with the element charge $e$, (c) $S$, (d) $S^2/\rho_{xx}$, (e) $\kappa$, and (f) $ZT$. The dotted and dashed lines are guides to the eyes. The absolute values of the resistivities are not so different at 300~K and 800~K. The mobilities are plotted in Fig.~\ref{fig8}(b). Since Hall-effect measurements were carried out only at 300~K and not at elevated temperatures, panel (b) contains only 300~K data. The suppression of $\rho_{xx}$ with $x$ is almost fully ascribed to the increase in $n$: An almost constant mobility is observed in this system. The thermopower at 300~K and 800~K given in Fig.~\ref{fig8} (c) decreases upon Ta doping, in agreement with the decrease in the resistivity. The difference $\Delta S = S_{\rm 800~K}-S_{\rm 300~K}$ in the absolute values for each sample amounts to around 100$\mu$V/K. The maximum of the power factor is clearly seen in Fig.~\ref{fig8} (d) between $n=0.02$ and 0.04 at both temperatures. The absolute values increase by a factor of three or four when going from 300~K to 800~K. At the same time the average thermal conductivity drops roughly to half of its room-temperature value, as depicted by the dotted and dashed lines in Fig.~\ref{fig8} (e). This reduction of the thermal conductivity should be attributed to an increased phonon-phonon scattering rate. The dependence of the FOM on $n$ shown in Fig.~\ref{fig8} (f) resembles that of the power factor and also exhibits a maximum in the intermediate Ta-doping range. The increase in $ZT$ between 300~K and 800~K amounts to more than one order of magnitude.
\begin{table*}[t]
\centering
\begin{ruledtabular}
\caption{Thermoelectric properties of  \WTS\ (from Ref.~\onlinecite{brixner63a}) and \WTST\ (this work). The actual charge-carrier concentration $n$ is presented in holes per W site, the resistivity in m$\Omega$cm, the thermopower in $\mu$V/K, the power factor in $\mu$W/K$^2$~cm, and the thermal conductivity in W/K~m; see text for details.}
\label{tab1}
\begin{tabular}{cccccccccccc}
\toprule
                                         &              & \multicolumn{5}{c}\WTS\                                                             &   \multicolumn{5}{c}\WTST\                                                       \\ \addlinespace[0.5em]
\cline{3-7}\cline{8-12}
\addlinespace[0.5em]
$n$                                   & $T$       & $\rho$ & $S$  & $S^2/\rho$  & $\kappa_{\rm tot}$ & $ZT$   & $\rho$ & $S$  & $S^2/\rho$  & $\kappa_{\rm tot}$ & $ZT$\\ \addlinespace[0.5em]\hline \addlinespace[0.5em]
\multirow{2}{*}{0.006} & 300~K  &      7.0  & 112  &  1.8             &   6.87                        & 0.01 & 23.0    & 194  &  1.6              & 4.11                          & 0.01           \\ \addlinespace[0.1em]
                                         & 800~K  &       6.9 &  335 & 16.3            &   2.56                        & 0.60  & 15.2    & 302  &  6.0              & 2.00                          & 0.24           \\ \addlinespace[0.1em]
\multirow{2}{*}{0.027} & 300~K  &      2.0  & 90    &  4.0             &   7.31                        & 0.05 &   5.1    & 105  &  2.2              & 3.10                          & 0.02          \\ \addlinespace[0.1em]
                                         & 800~K  &       3.1 & 166  &  8.9              &    2.23                      & 0.35  &  5.8    & 210  &  7.6              & 1.90                          & 0.32           \\ \addlinespace[0.1em]                  
\bottomrule
\end{tabular}
\end{ruledtabular}
\end{table*}

Finally, we compare the results on \WTST\ with those on \WTS\ reported in the literature. There is one paper focusing comprehensively on the thermoelectric properties in \WTS\ by Brixner.\cite{brixner63a} Additional transport data from the same group are summarized in Ref.~\onlinecite{hicks64a}. In Ref.~\onlinecite{brixner63a}, the temperature dependences of $\rho$, $S$, $\kappa$, and $ZT$ are shown for $x = 0.01$ and 0.03. The respective values for the FOM at 800~K are $ZT \approx 0.6$ for $x=0.01$ and $ZT \approx 0.35$ for $x=0.03$, exceeding the maximum $ZT$ value of 0.32 reported here for \WTST. The thermoelectric parameters for these samples are summarized for comparison in Table~\ref{tab1}. There are several differences between these systems: (i) For small $x$, the resistivity is smaller and the thermopower is larger in \WTS\ than in \WTST, leading to larger power factors. (ii) Due to the disorder-scattering-induced reduction of the phononic mean-free path, the thermal conductivity in the Se-substituted system with $y=0.4$ is clearly smaller at room temperature than the value of \WTS\ ($y=0$) reported in Ref.~\onlinecite{brixner63a}. Apparently, this merit of the Te codoping is overcompensated by the increase in the resistivity due to additional disorder in the anion sublattice, which increases the scattering rate of the $p$-type charge carriers. (iii) The thermal conductivity\cite{brixnercomment} at 800~K in \WTS\ is larger than 2~W/K m, but almost comparable to \WTST, and hence at high temperatures the FOM is larger in the ``pure'' diselenide system. 

\begin{table}[b]
\centering
\begin{ruledtabular}
\caption{Electronic \kel\ and phononic \kph\ contributions to the total thermal conductivities $\kappa_{\rm tot}=\kph+\kel$  in \WTS\ (from Ref.~\onlinecite{brixner63a}) and \WTST\ (this work). The actual charge-carrier concentration $n$ is presented in holes per W site and the thermal conductivity values in W/K~m; see text for details.}
\label{tab2}
\begin{tabular}{cccccccc}
\toprule
                                         &              & \multicolumn{3}{c}\WTS\                                                             &   \multicolumn{3}{c}\WTST\                                                       \\ \addlinespace[0.1em]
\cline{3-5}\cline{6-8}
$n$                                   & $T$       & $\kappa_{\rm el}$ & $\kappa_{\rm ph}$ & $\kappa_{\rm tot}$  & $\kappa_{\rm el}$ & $\kappa_{\rm ph}$ & $\kappa_{\rm tot}$ \\ \addlinespace[0.25em]\hline \addlinespace[0.75em]
\multirow{2}{*}{0.006} & 300~K  & 0.10                        & 6.77                         & 6.87                           & 0.03                        & 4.08                         & 4.11                         \\ \addlinespace[0.1em]
                                         & 800~K  & 0.28                        & 2.28                         & 2.56                           & 0.13                        & 1.87                         & 2.00                         \\ \addlinespace[0.1em]
\multirow{2}{*}{0.027} & 300~K  & 0.36                        & 6.95                         & 7.31                           & 0.14                        & 2.96                         & 3.10                         \\ \addlinespace[0.1em]
                                         & 800~K  & 0.64                        & 1.59                         & 2.23                           & 0.33                        & 1.57                         & 1.90                         \\ \addlinespace[0.1em]                  
\bottomrule
\end{tabular}
\end{ruledtabular}
\end{table}
To see whether the phononic contributions to the total thermal conductivity are different, we assume that the Wiedemann-Franz law holds in these systems, and calculate the contribution of the charge carriers via $\kel = L_0 (T/\rho)$ with the Lorenz number $L_0 = 2.44 \times 10^{-8}$~V$^2$/K$^2$ of the Drude-Sommerfeld model. By subtracting \kel\ from the total thermal conductivity \ktot, the phononic part \kph\ was calculated. In \WTST, the electronic thermal conductivity \kel\ increases linearly with $n$ (not shown) and amounts at room temperature to $\sim 0.03 - 0.3$~W/K~m and at 800~K to $\sim 0.1 - 0.5$~W/K~m. This is less than 22\% of the total thermal conductivity, and most of the thermal conductivity is attributed to the phononic contribution. From the data given in Ref.~\onlinecite{brixner63a}, we estimated the respective electronic and phononic thermal conductivities in \WTS. To readily compare these numbers to \WTST, one has to take into account the real charge-carrier concentration per W site as estimated from Hall-effect measurements. In Ref.~\onlinecite{brixner63a} the hole concentration is only given for $x=0.01$ which corresponds to 0.004 per W site. Following Ref.~\onlinecite{hicks64a}, the carrier concentration for $x = 0.03$ is 0.02 per W site. Using these numbers, the electronic and phononic contributions of the thermal conductivities from Ref.~\onlinecite{brixner63a} have to be compared with the respective values of our \WTST\ samples for nominal $x = 0.02$ ($n\approx 0.006$) and $x=0.035$ ($n\approx 0.027$); see Fig.~\ref{fig2} (c). Table~\ref{tab2} summarizes the different contributions for these two charge-carrier concentrations at 300~K and 800~K. In both systems, the thermal conductivity is dominated by the phononic contribution. However, the phononic thermal conductivity is smaller in \WTST\ than in \WTS, especially at 300~K. Hence both components of the thermal conductivity $\kappa = \kph + \kel$ are substantially reduced in \WTST. It might be promising to further change the Te concentration and see whether one can optimize the thermoelectric properties and exceed the FOM reported for \WTS.

\section{Summary}
\label{summ}
We present a comprehensive theoretical and experimental study on \WTST\ for small Ta concentrations $0\leq x \leq 0.06$. From band-structure calculations and an analysis of the specific heat, we find clear evidence that upon introducing Te into the Se sites in WSe$_2$, the band structure changes. The isotropic band at the $\Gamma$ point is lowered in energy while the anisotropic bands at the $K$ and $K^\prime$ points shift toward the Fermi level, leading to a change in the doped-hole character. This crossover was monitored in the electronic specific-heat coefficient which reflects the apparent change in the filling dependence of the density of states, namely, from filling charge carriers solely into the isotropic band at the $\Gamma$ point in \WTS\ toward filling into both the isotropic band at the $\Gamma$ point and the anisotropic bands at the $K$ and $K^\prime$ points in \WTST.

Thermal- and electronic-transport measurements up to 850~K on this system yield that the maximum thermoelectric figure of merit is about 0.3 at 850~K in the doping range $0.03 \leq x \leq 0.035$. In comparison to the \WTS\ system, Te doping was found to successfully suppress the thermal conductivity, especially around room temperature. However, this merit is overcompensated by an increased resistivity due to the additional disorder in the anion sublattice, which leads to stronger scattering of the hole carriers.


\section*{Acknowledgments}
We thank A.~Yamamoto and T.~Ideue for technical assistance as well as M.~S.~Bahramy for fruitful discussions.
This study was supported by the Funding Program for World-Leading Innovative R\&D on Science and Technology (FIRST Program) from JSPS.
MK is supported by a Grant-in-Aid for Young Scientists (B) from the Japan Society for the Promotion of Science (JSPS KAKENHI Grant No. 25800197).


\begin{thebibliography}{75}
\bibitem[{Sin()}]{SingleCrystcomment}
\bibinfo{howpublished}{{Single crystal growth of the undoped dichalcogenides by
  vapor transport has been
  reported.\cite{nitsche61a,brixner62a,kershaw67a,alhilli72a,agarwal85a,srivastava85a,hofmann86a}
  Halogen-transport agents were often used to change the intrinsic
  charge-carrier concentration and its sign. A few reports focused on vapor
  transport without the use of transport agents. We followed the recipe given
  in Ref.~\onlinecite{alhilli72a} and successfully prepared $\sim 10 \mu$m
  thick single crystalline sheets ($\sim 0.5$~cm$^2$) of pristine WSe$_2$, but
  could not obtain single-crystalline-doped materials.}}
\bibitem[{Ska()}]{Skappacomment}
\bibinfo{howpublished}{{Since for the two quantities ${S}_{xx}$ and
  $\kappa_{xx}$, the longitudinal $xx$ component is the only one reported in
  this paper, we will refer to them as ${S}$ and $\kappa$ for simplicity in the
  following.}}
\bibitem[{cpc()}]{cpcvcomment}
\bibinfo{howpublished}{{Strictly speaking, one should also include the
  difference between the constant-pressure specific heat $c_p$ measured here
  and the constant-volume specific heat $c_v$ used in theoretical derivations
  of the underlying formula which are used to estimate the high-temperature
  specific heat in the calculation of the thermal conductivity $\kappa = D \,
  c_p\,d_{_{\rm 300K}}$. Conventionally, this difference is neglected.
  Nevertheless we estimated it and found that the difference $c_p -
  c_v=(\alpha^2 / \beta /d_{_{\rm 300K}}) T$ with the volume thermal-expansion
  coefficient $\alpha$, the isothermal compressibility $\beta$, and the
  room-temperature density $d_{_{\rm 300K}}$ amounts to $\sim 1.8$\% of the
  total specific heat at $T= 800$~K reducing the maximum figure of merit by the
  same 1.8\%. Since this is still much smaller than the total error bar in the
  thermal-diffusivity measurement, it is neglected here. For the estimation,
  $\alpha$ was taken from Ref.~\onlinecite{brixner63a} and $\beta$ from
  Ref.~\onlinecite{selvi08a}.}}
\bibitem[{BSc()}]{BScomment}
\bibinfo{howpublished}{{We note that in literature, there are contradicting
  results found about the position of the highest valence-band maximum in
  momentum space; see, e.g., Refs.~\onlinecite{hyuan13a},
  \onlinecite{coehoorn87a}, and \onlinecite{dawson87a}, \onlinecite{sharma99a}.
  The origin seems to be the choice of the structural parameters, e.g., the
  W-Se layer distance $z$. Experimentally, it is also not clear where the
  valence-band maximum is located; see, e.g.,
  Refs.~\onlinecite{traving97a,finteis97a,finteis99a}.}}
\bibitem[{bri()}]{brixnercomment}
\bibinfo{howpublished}{{The unit of $\kappa$ in Table III in
  Ref.~\onlinecite{brixner63a} is wrong. It should read ``W/cm-deg''.}}
\end{thebibliography}

\end{document}